\documentclass{article}
\usepackage{spconf,amsmath,graphicx}
\usepackage[utf8]{inputenc}
\usepackage{amsfonts}
\usepackage{booktabs}
\usepackage{cite}

\def\OrigLoss{{\mathcal{L}_{\text{CTC}}}}
\def\InterLoss{{\mathcal{L}_{\text{InterCTC}}}}
\def\InterPos{{\left\lfloor L/2 \right\rfloor}}

\title{Intermediate Loss Regularization for CTC-based Speech Recognition}
\name{Jaesong Lee$^1$, Shinji Watanabe$^2$}
\address{$^1$Naver Corporation $^2$Johns Hopkins University}

\begin{document}
\ninept
\maketitle
\begin{abstract}
We present a simple and efficient auxiliary loss function for automatic speech recognition (ASR) based on the connectionist temporal classification (CTC) objective.
The proposed objective, an intermediate CTC loss, is attached to an intermediate layer in the CTC encoder network.
This intermediate CTC loss well regularizes CTC training and improves the performance requiring only small modification of the code and small and no overhead during training and inference, respectively.
In addition, we propose to combine this intermediate CTC loss with stochastic depth training, and apply this combination to a recently proposed Conformer network.
We evaluate the proposed method on various corpora, reaching word error rate (WER) 9.9\% on the WSJ corpus and character error rate (CER) 5.2\% on the AISHELL-1 corpus respectively, based on CTC greedy search without a language model.
Especially, the AISHELL-1 task is comparable to other state-of-the-art ASR systems based on auto-regressive decoder with beam search.
\end{abstract}
\section{Introduction}
\label{sec:intro}

End-to-end automatic speech recognition (ASR) has become a promising approach for the speech recognition community.
It simplifies model design, training, and decoding procedure compared to conventional approaches like hybrid systems using hidden Markov model (HMM).

However, the improvement comes with a computational cost:
many state-of-the-art ASR architectures employ attention-based deep encoder-decoder architecture~\cite{chorowski2015attention, 7472621, 8462506, Karita2019, gulati2020conformer},
which requires heavy computational cost and large model size.
Also, the decoder runs in an autoregressive fashion and requires sequential computation, i.e., the generation of an output token can be started only after the completion of the previous token.

Compared to the encoder-decoder modeling,
the Connectionist Temporal Classification (CTC)~\cite{graves2006connectionist}
does not require a separate decoder, thus allows designing more compact and fast models.
Also, CTC provides a greedy decoding algorithm for generating sentences in a fast and parallel way,
especially compared to autoregressive decoder of encoder-decoder models.

Although recent advances on architectural design~\cite{Pratap2020,quartznet} and pre-training method~\cite{wav2vec2} have improved the performance with CTC,
it is usually weaker than encoder-decoder models, often credited to its strong conditional independence assumption,
and overcoming the performance often requires external language models (LMs) and beam search algorithm \cite{miao2015eesen,ueno2018acoustic},
which demand extra computational costs and effectively makes the model an autoregressive one.
Therefore, it is important to improve CTC modeling to reduce overall computational overhead,
ideally without the help of LM and beam search.

There also has been a great interest on non-autoregressive speech recognition toward reaching the performance of autoregressive models~\cite{chen2019non,higuchi2020mask,chan2020imputer,fujita2020insertion,tian2020spike}, inspired by the success of non-autoregressive models in neural machine translation \cite{gu2018non,ghazvininejad2019mask, ma2019flowseq,Shu2020LaNMT}.
Non-autoregressive ASR would allow faster token generation compared to autoregressive ASR,
as the generation of a token does not directly depend on the previous token.
CTC itself can be viewed as an early instance of non-autoregressive ASR,
and recently proposed methods, Mask CTC~\cite{higuchi2020mask} and Imputer~\cite{chan2020imputer},
use CTC as a part of non-autoregressive modeling:
they first generate initial output from CTC, then refine it via the other network.
Therefore, improving CTC is also important for improving non-autoregressive methods in general.

In this work, we show the performance of CTC can be improved with a proposed auxiliary task.
The proposed task, named intermediate CTC loss, is constructed by first obtaining the intermediate representation of the model then computing its corresponding CTC loss.
The model is trained with the original CTC loss in conjunction with the proposed loss, with a very small computational overhead.
During inference, the usual CTC decoding algorithm is used, thus there is no overhead.

We show the proposed method can improve Transformer~\cite{NIPS2017_7181} with various depths,
and also Conformer~\cite{gulati2020conformer}, recently proposed architecture combining self-attention and convolution layers.
Also, we show the method can be combined with the other regularization method, stochastic depth~\cite{huang2016deep, Pham2019}, for further enhancement.

The contributions of this paper are as follows:
\begin{itemize}
\setlength{\itemsep}{0pt}\setlength{\parskip}{0pt}\setlength{\parsep}{0pt}
    \item
    We present a simple yet efficient auxiliary loss, called intermediate CTC loss,
    for improving performance of CTC ASR network.
    \item We combine the intermediate CTC loss and stochastic depth regularization
    to achieve better performance than using only one of them.
    \item We show application to the Conformer encoder, recently proposed architecture.
    We show the proposed method is also effective for Conformer.
    \item We achieve comparable to state-of-the-art results,
    specifically word error rate (WER) 9.9\% on Wall Street Journal (WSJ) and
    character error rate (CER) 5.2\% on AISHELL-1,
    using CTC modeling and greedy decoding only.
\end{itemize}

\section{Architecture}
\label{sec:architecture}

We consider a multi-layer architecture with the CTC loss function.
For given input $x_0 \in \mathbb{R}^{T \times D}$ of length $T$ and dimension $D$,
the encoder consists of $L$ layers as follows:

\begin{equation}
\label{equation:encoder}
x_l = \textbf{EncoderLayer}_l(x_{l-1}),
\end{equation}
where $\textbf{EncoderLayer}_l$ is the $l$-th layer of the network explained at Section~\ref{sec:encoder}.

\subsection{Connectionist Temporal Classification}

CTC~\cite{graves2006connectionist} computes the likelihood of target sequence $y$
by considering all possible alignments for the label and the input length $T$.
For the encoder output $x_L$ and target sequence $y$, the likelihood is defined as:
\begin{equation}
\label{equation:ctc_prob}
P_\mathsf{CTC}(y|x_L) := \sum_{a \in \beta^{-1}(y)} P(a|x_L)
\end{equation}
where $\beta^{-1}(y)$ is the set of alignment $a$ of length $T$ compatible to $y$ including the special blank token.
The alignment probability $P(a|x_L)$ is factorized with the following conditional independence assumption:
\begin{equation}
\label{equation:alignment_probability}
P(a|x_L) = \prod_{t} P(a[t]|x_L[t])
\end{equation}
where $a[t]$ and $x_L[t]$ denote the $t$-th symbol of $a$ and the $t$-th representation vector of $x_L$, respectively.

At training time, we minimize the negative log-likelihood induced by CTC by using $P_\mathsf{CTC}(y|x_L)$ in Eq.~\eqref{equation:ctc_prob}:
\begin{equation}
    \OrigLoss := - \log P_\mathsf{CTC}(y|x_L).
\end{equation}
At test time, we use greedy search to find the most probable alignment for fast inference.

\subsection{Encoder}
\label{sec:encoder}

We use two encoder architectures:
Transformer~\cite{NIPS2017_7181} and Conformer~\cite{gulati2020conformer}.
Transformer uses self-attention ($\text{SelfAttention}(\cdot)$ shown in Eq.~\eqref{equation:transformer}) for learning global representation,
and layer normalization~\cite{ba2016layer} and residual connection~\cite{7780459} for stabilizing learning.

With Transformer, \textbf{EncoderLayer}$(\cdot)$ in Eq.~\eqref{equation:encoder} consists of:
\begin{align}
\label{equation:transformer}
x^\text{MHA}_l &= \text{SelfAttention}(x_{l-1}) + x_{l-1}, \\
x_l &= \text{FFN}(x^\text{MHA}_l) + x^\text{MHA}_l,
\end{align}
where $\text{FFN}(\cdot)$ denotes the feed forward layers.

Conformer combines Transformer and convolution neural layers for efficient learning of both global and local representations.

With Conformer, \textbf{EncoderLayer}$(\cdot)$ in Eq.~\eqref{equation:encoder} consists of:

\begin{align}
x^\text{FFN}_l &= \frac{1}{2} \text{FFN}(x_{l-1}) + x_{l-1} \\
x^\text{MHA}_l &= \text{SelfAttention}(x^\text{FFN}_l) + x^\text{FFN}_l \\
x^\text{Conv}_l &= \text{Convolution}(x^\text{MHA}_l) + x^\text{MHA}_l \\
x_l &= \text{LayerNorm}(\frac{1}{2} \text{FFN}(x^\text{Conv}_l) + x^\text{Conv}_l).
\end{align}

\subsection{Stochastic Depth}
\label{sec:stochastic_depth}

Stochastic depth~\cite{huang2016deep, Pham2019} is a regularization technique for residual network.
It helps training of very deep networks by randomly skipping some layers. It can be viewed as training an ensemble of $2^L$ sub-models, induced by removing some layers of the model.

Consider \textbf{EncoderLayer}$(\cdot)$ in Eq.~\eqref{equation:encoder}  with residual connection:
\begin{equation}
\label{equation:encoder_residual}
x_l = x_{l-1} + f_l(x_{l-1})
\end{equation}
for some layer $f_l(\cdot)$.

Let $b_l$ be a Bernoulli random variable which takes value 1 with probability $p_l$.
During training, the layer is computed as:
\begin{equation}
\label{equation:stochastic_depth}
x_l = \begin{cases}
x_{l-1} & \text{if } b_l = 0, \\
x_{l-1} + \frac{1}{p_l} \cdot f_l(x_{l-1}) & \text{otherwise.}\\
\end{cases}
\end{equation}

Thus, with probability $1 - p_l$, the layer skips the $f_l(x_{l-1})$ part.
The denominator $\frac{1}{p_1}$ ensures the expectation matches the Eq.~\eqref{equation:encoder}.
During testing, we do not skip the layers and use Eq.~\eqref{equation:encoder_residual}.

The per-layer survival probability is given as $p_l = 1 - \frac{l}{L}(1 - p_L)$ with hyper-parameter $p_L$.
This assigns higher skipping probability to higher layers,
as skipping lower layers may harm the overall performance~\cite{huang2016deep}.
We use $p_L = 0.7$ for all experiments.

\section{Intermediate CTC Loss}
\label{sec:intermediate_ctc}

The stochastic depth aims to improve training of multi-layer network using a stochastic ensemble approach,
but the experiments show the improvement only comes with sufficiently deep networks,
e.g. with 24 or more layers~\cite{Pham2019}.

We hypothesize that while the stochastic depth is effective for regularizing higher layers,
it is not effective for regularizing lower layers,
due to the its ensemble strategy.
As each layer has own random variable for skipping,
the probability of skipping all high layers is very low.
Therefore, for most cases, the lower layers may rely on the remaining higher layers rather than learn regularized representation by themselves.

In this context, we propose to skip the higher layers as a whole.
We choose a layer, called ``intermediate layer'', and
induce a sub-model by skipping all layers after the intermediate layer.
The sub-model relies on the lower layers rather than higher layers,
thus training the sub-model would regularize the lower part of the full model.

For the position of the intermediate layer,
this paper mainly uses $\InterPos$,
as it seems a safe choice between lower and higher layers.
We later discuss other choices at Section~\ref{sec:position_variants}.

As the sub-model and the full model share lower structure, it is possible to denote the output of the sub-model as $x_\InterPos$, the intermediate representation of the full-model.
Like the full-model, we use a CTC loss for the sub-model:
\begin{equation}
    \label{equation:inter_loss}
    \InterLoss := - \log P_\mathsf{CTC}(y|x_\InterPos).
\end{equation}
Then we note that the sub-model representation $x_\InterPos$ is naturally obtained
when we compute the full model.
Thus, after computing the CTC loss of the full model,
we can compute the CTC loss of the sub-model with a very small overhead.
The proposed training objective is the weighted sum of the two losses:
\begin{equation}
    \label{equation:total_loss}
    \mathcal{L} := (1-w) \OrigLoss + w \InterLoss,
\end{equation}
where we use $w = 0.3$ for all experiments.

During testing, we do not use the intermediate prediction and only use the final representation $x_L$ for decoding.

The intermediate loss can also be used jointly with stochastic depth.
We expect the intermediate loss regularizes the lower layers,
and the stochastic depth regularizes the higher layers,
thus combining them further improves the whole model.
We show the empirical result at Section~\ref{sec:experiments}.

\subsection{Position variants}
\label{sec:position_variants}

We also consider different sub-model configurations and investigate their effects.
We consider the following variants:

\begin{itemize}
\setlength{\itemsep}{0pt}\setlength{\parskip}{0pt}\setlength{\parsep}{0pt}
\item {\bf Lower than the middle}.
Depending on the number of layers $L$, the optimal ratio of lower layers to higher layers may differ.
To find the effect of position of the intermediate loss,
we consider lower position than middle, e.g., $\left\lfloor L/4 \right\rfloor$, for the sub-model.

\item {\bf Multiple sub-models}.
We consider multiple sub-models rather than only one.
For the number of sub-models $K$, we compute the following loss:
\begin{equation}
- \frac{1}{K}
    \sum_{k=1}^{K}
    P_\mathsf{CTC}(y|x_{\left\lfloor \frac{k L}{K + 1} \right\rfloor}).
\end{equation}
For $K  = 1$, the loss corresponds to Eq.~\eqref{equation:inter_loss}.

\item {\bf Random position}.
We also consider randomly choosing sub-model among multiple models.
We introduce a uniform random variable $u$ with range from $\InterPos$ to $L - 1$,
and choose $u$-th layer for the intermediate representation.
\end{itemize}

We show the experimental results at Section~\ref{sec:exp_variants}.

\subsection{Stochastic variant of Intermediate Loss}
\label{sec:stoch_variant}

In Eq.~\eqref{equation:total_loss}, we compute the weighted sum of the two sub-models.
Instead, we may compute the stochastic variant of the loss, like stochastic depth, as follows.
Let $b$ a Bernoulli random variable which takes value 1 with probability $w$.
the stochastic intermediate CTC objective is:
\begin{equation}
    \label{equation:stoch_inter_ctc}
    \mathcal{L}' := \begin{cases}
        \OrigLoss & \text{if } b = 0,\\
        \InterLoss & \text{otherwise}.
    \end{cases}
\end{equation}
The loss coincides with Eq.~\eqref{equation:total_loss} in expectation.

We argue the deterministic version is better than stochastic one for gradient-based learning
even if they have same expected value.
For the stochastic variant,
the loss and its gradient only have access to $\InterLoss$ if $b = 1$,
and the model may forget features useful for $\OrigLoss$ but not for $\InterLoss$.
On the other hand, the deterministic variant always computes two losses at the same time,
therefore, the risk of forgetting features is low.

At Section~\ref{sec:exp_variants},
we experimentally show while the stochastic variant also improves the model,
it is not so effective as the deterministic one.

\subsection{Application to other non-autoregressive ASR: Mask CTC}
\label{sec:mask_ctc}

Mask CTC~\cite{higuchi2020mask} consists of an encoder, a CTC layer on top of the encoder,
and a conditional masked language model (CMLM) \cite{ghazvininejad2019mask}.

During decoding, the model first generates initial hypotheses from the CTC layer,
and replaces any token of low probability (below a given threshold) with special token \texttt{<MASK>}.
The CMLM predicts the token of masked position given the masked hypothesis.

During training, the target $y$ is randomly masked and fed to CMLM.
The CMLM predicts the token of masked position for the masked input.
Let $y_\text{obs}$ the masked input and $y_\text{mask}$ the prediction for the mask.
The training objective is:
\begin{multline}
- w_\mathsf{CTC} \log P_\mathsf{CTC}(y|x_L)
- (1 - w_\mathsf{CTC}) \log P_\mathsf{CMLM}(y_\mathsf{mask}|y_\mathsf{obs}, x_L)
\end{multline}
with hyper-parameter $w_\mathsf{CTC}$.

As the initial hypothesis is predicted from the CTC layer, its performance is crucial for the overall performance.
We aim to improve the CTC layer using the proposed intermediate loss.
We take the intermediate output $x_\InterPos$ from the encoder and compute the intermediate CTC probability $P_\mathsf{CTC}(y|x_\InterPos)$.
The extended training objective is:
\begin{multline}
- w_\mathsf{CTC} \log P_\mathsf{CTC}(y|x_L)
- w_\mathsf{InterCTC} \log P_\mathsf{CTC}(y|x_\InterPos) \\
- (1 - w_\mathsf{CTC} - w_\mathsf{InterCTC}) \log P_\mathsf{CMLM}(y_\mathsf{mask}|y_\mathsf{obs}, x_L).
\end{multline}
We present the experimental result for Mask CTC at Section~\ref{sec:exp_mask_ctc}.

\subsection{Related work}
\label{sec:comparison_hctc}

Hierarchical CTC~\cite{10.5555/1625275.1625400,DBLP:journals/corr/abs-1807-06234,Toshniwal2017} (HCTC) introduced an auxiliary CTC task based on the assumption that different layers learn different level of abstraction.
While HCTC looks similar to intermediate loss,
it requires additional labeling effort (e.g., phoneme)
or various tokenization (e.g., sub-word for high-level and character for low-level),
which may not be applicable for certain cases,
e.g., when the character-based tokenization is the best effort for the data like Mandarin and Japanese \cite{watanabe2018espnet}.
In contrast, intermediate CTC is based on the sub-model regularization, therefore it does not require additional low-level labels,
and it is natural to combine intermediate CTC with stochastic depth.

\cite{9052964} and
\cite{liu2020improving} introduced additional networks to train the intermediate layer of the encoder for CTC and RNN-Transducer~\cite{graves2012sequence} respectively.
We note that intermediate CTC does not require additional network,
and has very little overhead at the training time, in contrast to~\cite{liu2020improving}, due to the structure of CTC architecture.

\section{Experiments}
\label{sec:experiments}

We evaluate the performance of intermediate CTC loss on the three corpora:
Wall Street Journal (WSJ)~\cite{paul1992design} (English, 81 hours),
TED-LIUM2~\cite{rousseau-etal-2014-enhancing} (English, 207 hours),
and AISHELL-1~\cite{bu2017aishell} (Chinese, 170 hours).
We use ESPnet~\cite{watanabe2018espnet} for all experiments.
We use 80-dimensional log-mel feature and 3-dimensional pitch feature for the input,
and apply SpecAugment~\cite{park2019specaugment} during training.
For WSJ and AISHELL-1, we tokenize label sentences as characters.
For TED-LIUM2, we tokenize label sentences as sub-words with sentencepiece~\cite{kudo-richardson-2018-sentencepiece}.

For WSJ, the model is trained for 100 epochs.
For TED-LIUM2 and AISHELL-1, the model is trained for 50 epochs.
After training, the model parameter is obtained by averaging models from last 10 epochs.
Note that we do not use any external language models (LMs) or beam search, and only use greedy decoding for CTC.
Thus, all experiments are based on the \textit{non-autoregressive} setup in order to keep the benefit of fast and parallel inference of CTC.


\subsection{Results}
\label{sec:experimental_results}

We show the experimental results for Transformer and Conformer architectures.
For each architecture, we compare four regularization configurations:

\begin{itemize}
\setlength{\itemsep}{0pt}\setlength{\parskip}{0pt}\setlength{\parsep}{0pt}
\item Baseline (no regularization)
\item Intermediate CTC (``InterCTC'')
\item Stochastic depth (``StochDepth'')
\item Intermediate CTC + Stochastic depth (``both'')
\end{itemize}

\begin{table}[t]
\centering
\caption{Word error rates (WERs) and character error rates (CERs) for Transformer.
See section \ref{sec:experimental_results} for details.}
\label{table:transformer}

\scalebox{0.9}{

\begin{tabular}{l | cc | cc | cc}
\toprule
&
\multicolumn{2}{c}{WSJ} &
\multicolumn{2}{| c}{TED-LIUM2} &
\multicolumn{2}{| c}{AISHELL-1} \\
& \multicolumn{2}{c}{(WER)}
& \multicolumn{2}{| c}{(WER)}
& \multicolumn{2}{| c}{(CER)} \\

& dev93 & eval92 & dev & test & dev & test \\

\midrule

\bf{12-layer} & 20.1 & 16.5 & 14.8 & 14.0 & 5.8 & 6.3 \\
\quad + InterCTC & 17.5 & 13.6 & 13.3 & 12.3 & 5.7 & 6.2 \\
\quad + StochDepth & 19.8 & 16.2 & 13.8 & 13.1 & 5.9 & 6.4 \\
\quad + both & 16.8 & 13.7 & 13.2 & 12.1 & 5.7 & 6.1 \\

\midrule

\bf{24-layer} & 17.8 & 13.9 & 12.6 & 12.2 & 5.4 & 5.9 \\
\quad + InterCTC & 15.3 & 12.4 & 11.5 & 10.6 & 5.1 & 5.6 \\
\quad + StochDepth & 16.3 & 12.7 & 11.9 & 11.2 & 5.2 & 5.7 \\
\quad + both & 14.9 & \textbf{11.8} & 10.9 & 10.2 & 5.2 & 5.5 \\

\midrule

\bf{48-layer} & 16.6 & 13.8 & 11.6 & 10.9 & 5.1 & 5.7 \\
\quad + InterCTC & 14.9 & 12.6 & 10.7 & 10.3 & 5.1 & 5.5 \\
\quad + StochDepth & 15.6 & 12.9 & 11.0 & 10.2 & 5.0 & 5.4 \\
\quad + both & \textbf{14.2} & \textbf{11.8} & \textbf{10.3} & \textbf{9.9} & \textbf{4.9} & \textbf{5.3} \\

\bottomrule
\end{tabular}

}
\vspace{-3mm}
\end{table}

\begin{table}[t]
\centering
\caption{Word error rates (WERs)
and character error rates (CERs)
for Conformer.
See section \ref{sec:experimental_results} for details.
}
\label{table:conformer}

\scalebox{0.9}{

\begin{tabular}{l | cc | cc
| cc 
}
\toprule
& \multicolumn{2}{c}{WSJ}
& \multicolumn{2}{| c}{TED-LIUM2}
& \multicolumn{2}{| c}{AISHELL-1}
\\
& \multicolumn{2}{c}{(WER)}
& \multicolumn{2}{| c}{(WER)}
& \multicolumn{2}{| c}{(CER)}

\\
& dev93 & eval92
& dev & test
& dev & test
\\

\midrule


\textbf{12-layer}
& 15.2 & 12.4 & 10.5 &9.8
& 5.4 & 6.0
\\
\quad + InterCTC
& 13.4 & 10.8 & \textbf{9.7} & \textbf{9.1}
& 5.1 & 5.6
\\

\quad + StochDepth & 13.1 & 10.8 & 11.1 & 10.7 & 5.2 & 5.8 \\

\quad + both
& \textbf{12.0} & \textbf{9.9} & 10.8 & 9.9
& \textbf{4.7} & \textbf{5.2}
\\

\bottomrule
\end{tabular}
}
\vspace{-3mm}
\end{table}

For Transformer, we use 12-layer, 24-layer and 48-layer models.
Table~\ref{table:transformer} shows the word error rates (WERs) for WSJ and TED-LIUM2,
and character error rates (CERs) for AISHELL-1.

For all of the experiment, intermediate CTC gives an improvement over the baseline model.
Stochastic depth improves 24-layer and 48-layer models, but does not improve 12-layer models well for WSJ and AISHELL-1.
Using both the intermediate loss and the stochastic depth gives better result than using only one of them.
Thus, we conclude the two methods have complimentary effects.

Additionally, we apply intermediate CTC to 6-layer Transformer for WSJ, and get WER improvement from 21.1\% to 18.3\%. This suggests the intermediate CTC is still beneficial for smaller networks.

For Conformer, we use 12-layer model.
The results are at Table~\ref{table:conformer}.
Again, intermediate CTC gives consistent improvement over baseline.
Stochastic depth gives improvement for WSJ and AISHELL-1, but does not give improvement for TED-LIUM2.

The combination of intermediate loss and stochastic depth achieves WER \textbf{9.9\%} for WSJ and CER \textbf{5.2\%} for AISHELL-1.
For WSJ, it outperforms the previously published non-autoregressive results~\cite{higuchi2020mask,chan2020imputer,chi2020align},
and is close to the state-of-the-art autoregressive result (9.3\%)~\cite{sabour2018optimal}.
Also, for AISHELL-1, it outperforms Transformer-based encoder-decoder models~\cite{karita2019comparative,gao2020sanm},
and is close to the state-of-the-art autoregressive result (5.1\%)~\cite{zhou2020selfandmixed}.
Note that the referred state-of-the-art results use an autoregressive decoder and \cite{zhou2020selfandmixed} also uses an external LM.
On the other hand, our result is solely based on CTC with greedy decoding, without LM or beam search.

\subsection{Study on Intermediate Loss design}
\label{sec:exp_variants}

\begin{table}[t]
\centering

\caption{Word error rates (WERs) of the intermediate loss variants for WSJ.
See Section~\ref{sec:exp_variants} for details.}
\label{table:variants}
\scalebox{0.8}{
\begin{tabular}{l l|c c}
\toprule
& & dev93 & eval92 \\
\midrule
\bf{12-layer}
& Default & 17.5 & 13.6 \\
& Random & 17.4 & 14.3 \\
& Stochastic & 19.0 & 15.0 \\

\midrule
\bf{24-layer}
& Default & 15.3 & 12.4 \\
& Lower & 15.8 & 12.9 \\
& Multiple & 15.1 & 12.0 \\
& Random & 15.4 & 12.4 \\

\midrule
\bf{48-layer}
& Default & 14.9 & 12.6 \\
& Multiple & 15.4 & 12.1 \\
& Random & 14.7 & 12.0 \\
\bottomrule
\end{tabular}
}
\vspace{-4mm}
\end{table}

To compare the proposed intermediate loss to the position variants (Section~\ref{sec:position_variants})
and the stochastic variant (Section~\ref{sec:stoch_variant}),
we conduct additional experiments for WSJ corpus.
The result is at Table~\ref{table:variants}.
We conduct the following experiments:

\begin{itemize}
\setlength{\itemsep}{0pt}\setlength{\parskip}{0pt}\setlength{\parsep}{0pt}
    \item
    {\bf Lower position}.
    We conduct this variant for the 24-layer model,
    which is sufficiently deep to consider a lower position. We used 6th layer for the experiment.
    Despite the deep network, the variant performs slightly worse than the default.
    \item
    {\bf Multiple positions}.
    We conduct this variant for the 24-layer and the 48-layers,
    which are very deep and more sub-models may help.
    We use $K = 3$ for 24-layer and
    $K = 7$ for 48-layer,
    to select all layer positions of power of 6.
    It gives a small improvement for the 24-layer,
    but gives a mixed result for the 48-layer.
    \item
    {\bf Random position}.
    We conduct this variant for all models.
    The result is mixed:
    it gives no improvement for the 12-layer and the 24-layer,
    although a small improvement for 48-layer.
    \item
    {\bf Stochastic variant}.
    We conduct this variant for 12-layer model.
    As discussed in Section~\ref{sec:stoch_variant},
    the stochastic variant is worse than the deterministic one,
    although it is still better than no regularization.
\end{itemize}

From the experimental results, we conclude that the proposed design is a simple yet reasonable choice among the variants.

\subsection{Application to other non-autoregressive ASR}
\label{sec:exp_mask_ctc}

\begin{table}[t]
\centering

\caption{Word error rates (WERs) of Mask CTC-based non-autoregressive ASR for WSJ.
See Section~\ref{sec:exp_mask_ctc} for details.}
\label{table:mask_ctc}
\scalebox{0.8}{
\begin{tabular}{l|l|c c}
\toprule
& threshold & dev93 & eval92 \\
\midrule

\textbf{12enc-6dec} & 0.0 & 16.5 & 13.5 \\
& 0.999 & 15.7 & 12.9 \\
\midrule
+ InterCTC & 0.0 & 14.4 & 11.6 \\
& 0.999 & \textbf{14.1} & \textbf{11.3} \\
\midrule
Mask CTC~\cite{higuchi2020mask} & 0.999 & 15.4 & 12.1 \\
Align-Refine~\cite{chi2020align} & - & 13.7 & 11.4 \\
\bottomrule
\end{tabular}
}
\vspace{-4mm}
\end{table}
We present an experimental result of Mask CTC-based non autoregressive ASR and intermediate loss, as described at Section~\ref{sec:mask_ctc}.
The WSJ corpus is used for the experiment.
We use $w_\text{CTC} = 0.3$,
and for intermediate CTC variant, we also use
$w_\text{InterCTC} = 0.3$.
We use a Transformer model with 12-layer encoder and 6-layer decoder.
The model is trained for 500 epochs and parameters of last 60 epochs are averaged.

Table~\ref{table:mask_ctc} shows the WERs for Mask CTC.
The second column indicates the threshold of probability for CTC prediction; Mask CTC uses 0.999 by default. If threshold is 0.0, the model does not use the decoder and just treats the CTC result as the final prediction.
We see the intermediate CTC improves the performance of CTC prediction, from 13.5\% to 11.6\%.
We also see the improvement of CTC leads the overall improvement of Mask CTC,
as the WER reduced from 12.9\% to 11.3\%.
It is also lower than Align-Refine~\cite{chi2020align} (11.4\%)
which improves Mask CTC by modifying the role of CMLM.
This shows the intermediate loss helps the training of Mask CTC.

\section{Conclusion}

We present intermediate CTC loss, an auxiliary task for improving CTC-based speech recognition.
The proposed loss is easy to implement, has small overhead at training time and no overhead at test time.
We empirically show the intermediate CTC loss improves Transformer and Conformer architectures,
and combining the loss with stochastic depth further improves training,
reaching word error rate (WER) 9.9\% on WSJ and
character error rate (CER) 5.2\% on AISHELL-1,
without an autoregressive decoder or external language model.

\bibliographystyle{IEEEbib}
\bibliography{refs}

\begin{thebibliography}{10}

\bibitem{chorowski2015attention}
Jan~K Chorowski et~al.,
\newblock ``Attention-based models for speech recognition,''
\newblock in {\em Proc. NeurIPS}, 2015.

\bibitem{7472621}
W.~{Chan} et~al.,
\newblock ``Listen, attend and spell: A neural network for large vocabulary
  conversational speech recognition,''
\newblock in {\em Proc. ICASSP}, 2016.

\bibitem{8462506}
L.~{Dong}, S.~{Xu}, and B.~{Xu},
\newblock ``Speech-transformer: A no-recurrence sequence-to-sequence model for
  speech recognition,''
\newblock in {\em Proc. ICASSP}, 2018.

\bibitem{Karita2019}
Shigeki Karita et~al.,
\newblock ``{Improving Transformer-Based End-to-End Speech Recognition with
  Connectionist Temporal Classification and Language Model Integration},''
\newblock in {\em Proc. Interspeech}, 2019.

\bibitem{gulati2020conformer}
Anmol Gulati et~al.,
\newblock ``Conformer: Convolution-augmented transformer for speech
  recognition,''
\newblock in {\em Proc. Interspeech}, 2020.

\bibitem{graves2006connectionist}
Alex Graves et~al.,
\newblock ``Connectionist temporal classification: labelling unsegmented
  sequence data with recurrent neural networks,''
\newblock in {\em Proc. ICML}, 2006.

\bibitem{Pratap2020}
Vineel Pratap et~al.,
\newblock ``Scaling up online speech recognition using convnets,''
\newblock in {\em Proc. Interspeech}, 2020.

\bibitem{quartznet}
S.~{Kriman} et~al.,
\newblock ``Quartznet: Deep automatic speech recognition with 1d time-channel
  separable convolutions,''
\newblock in {\em Proc. ICASSP}, 2020.

\bibitem{wav2vec2}
Alexei Baevski et~al.,
\newblock ``wav2vec 2.0: A framework for self-supervisedlearning of speech
  representations,''
\newblock in {\em Proc. NeurIPS}, 2020.

\bibitem{miao2015eesen}
Yajie Miao, Mohammad Gowayyed, and Florian Metze,
\newblock ``{EESEN}: End-to-end speech recognition using deep rnn models and
  wfst-based decoding,''
\newblock in {\em Proc. ASRU}, 2015.

\bibitem{ueno2018acoustic}
Sei Ueno et~al.,
\newblock ``Acoustic-to-word attention-based model complemented with
  character-level ctc-based model,''
\newblock in {\em Proc. ICASSP}, 2018.

\bibitem{chen2019non}
Nanxin Chen et~al.,
\newblock ``Listen and fill in the missing letters: Non-autoregressive
  transformer for speech recognition,'' 2020.

\bibitem{higuchi2020mask}
Yosuke Higuchi et~al.,
\newblock ``Mask ctc: Non-autoregressive end-to-end asr with ctc and mask
  predict,''
\newblock in {\em Proc. Interspeech}, 2020.

\bibitem{chan2020imputer}
William Chan et~al.,
\newblock ``Imputer: Sequence modelling via imputation and dynamic
  programming,''
\newblock in {\em Proc. ICML}, 2020.

\bibitem{fujita2020insertion}
Yuya Fujita et~al.,
\newblock ``Insertion-based modeling for end-to-end automatic speech
  recognition,''
\newblock in {\em Proc. Interspeech}, 2020.

\bibitem{tian2020spike}
Zhengkun Tian et~al.,
\newblock ``Spike-triggered non-autoregressive transformer for end-to-end
  speech recognition,''
\newblock in {\em Proc. Interspeech}, 2020.

\bibitem{gu2018non}
Jiatao Gu et~al.,
\newblock ``Non-autoregressive neural machine translation,''
\newblock in {\em Proc. ICLR}, 2018.

\bibitem{ghazvininejad2019mask}
Marjan Ghazvininejad et~al.,
\newblock ``Mask-predict: Parallel decoding of conditional masked language
  models,''
\newblock in {\em Proc. EMNLP-IJCNLP}, 2019.

\bibitem{ma2019flowseq}
Xuezhe Ma et~al.,
\newblock ``Flowseq: Non-autoregressive conditional sequence generation with
  generative flow,''
\newblock in {\em Proc. EMNLP-IJCNLP}, 2019.

\bibitem{Shu2020LaNMT}
Raphael Shu et~al.,
\newblock ``Latent-variable non-autoregressive neural machine translation with
  deterministic inference using a delta posterior,''
\newblock in {\em Proc. AAAI}, 2020.

\bibitem{NIPS2017_7181}
Ashish Vaswani et~al.,
\newblock ``Attention is all you need,''
\newblock in {\em Proc. NeurIPS}, 2017.

\bibitem{huang2016deep}
Gao Huang et~al.,
\newblock ``Deep networks with stochastic depth,''
\newblock in {\em Proc. ECCV}, 2016.

\bibitem{Pham2019}
Ngoc-Quan Pham et~al.,
\newblock ``{Very Deep Self-Attention Networks for End-to-End Speech
  Recognition},''
\newblock in {\em Proc. Interspeech}, 2019.

\bibitem{ba2016layer}
Jimmy~Lei Ba, Jamie~Ryan Kiros, and Geoffrey~E. Hinton,
\newblock ``Layer normalization,'' 2016.

\bibitem{7780459}
K.~{He} et~al.,
\newblock ``Deep residual learning for image recognition,''
\newblock in {\em Proc. CVPR}, 2016.

\bibitem{10.5555/1625275.1625400}
Santiago Fern\'{a}ndez, Alex Graves, and J\"{u}rgen Schmidhuber,
\newblock ``Sequence labelling in structured domains with hierarchical
  recurrent neural networks,''
\newblock in {\em Proc. IJCAI}, 2007.

\bibitem{DBLP:journals/corr/abs-1807-06234}
Kalpesh Krishna, Shubham Toshniwal, and Karen Livescu,
\newblock ``Hierarchical multitask learning for ctc-based speech recognition,''
  2019.

\bibitem{Toshniwal2017}
Shubham Toshniwal et~al.,
\newblock ``Multitask learning with low-level auxiliary tasks for
  encoder-decoder based speech recognition,''
\newblock in {\em Proc. Interspeech}, 2017.

\bibitem{watanabe2018espnet}
Shinji Watanabe et~al.,
\newblock ``{ESPnet}: End-to-end speech processing toolkit,''
\newblock in {\em Proc. Interspeech}, 2018.

\bibitem{9052964}
A.~{Tjandra} et~al.,
\newblock ``Deja-vu: Double feature presentation and iterated loss in deep
  transformer networks,''
\newblock in {\em Proc. ICASSP}, 2020.

\bibitem{liu2020improving}
Chunxi Liu et~al.,
\newblock ``Improving rnn transducer based asr with auxiliary tasks,''
\newblock in {\em Proc. SLT}, 2020.

\bibitem{graves2012sequence}
Alex Graves,
\newblock ``Sequence transduction with recurrent neural networks,'' 2012.

\bibitem{paul1992design}
Douglas~B Paul and Janet~M Baker,
\newblock ``The design for the wall street journal-based {CSR} corpus,''
\newblock in {\em Proc. Workshop on Speech and Natural Language}, 1992.

\bibitem{rousseau-etal-2014-enhancing}
Anthony Rousseau, Paul Del{\'e}glise, and Yannick Est{\`e}ve,
\newblock ``Enhancing the {TED}-{LIUM} corpus with selected data for language
  modeling and more {TED} talks,''
\newblock in {\em Proc. {LREC}}, May 2014.

\bibitem{bu2017aishell}
Hui Bu et~al.,
\newblock ``Aishell-1: An open-source mandarin speech corpus and a speech
  recognition baseline,''
\newblock in {\em Proc. O-COCOSDA}, 2017.

\bibitem{park2019specaugment}
Daniel~S Park et~al.,
\newblock ``{SpecAugment}: A simple data augmentation method for automatic
  speech recognition,''
\newblock in {\em Proc. Interspeech}, 2019.

\bibitem{kudo-richardson-2018-sentencepiece}
Taku Kudo and John Richardson,
\newblock ``{S}entence{P}iece: A simple and language independent subword
  tokenizer and detokenizer for neural text processing,''
\newblock in {\em Proc. EMNLP: System Demonstrations}, Nov. 2018.

\bibitem{chi2020align}
Ethan~A. Chi, Julian Salazar, and Katrin Kirchhoff,
\newblock ``Align-refine: Non-autoregressive speech recognition via iterative
  realignment,'' 2020.

\bibitem{sabour2018optimal}
Sara Sabour, William Chan, and Mohammad Norouzi,
\newblock ``Optimal completion distillation for sequence learning,''
\newblock in {\em Proc. ICLR}, 2019.

\bibitem{karita2019comparative}
S.~{Karita} et~al.,
\newblock ``A comparative study on transformer vs rnn in speech applications,''
\newblock in {\em Proc. ASRU}, 2019.

\bibitem{gao2020sanm}
Zhifu Gao et~al.,
\newblock ``San-m: Memory equipped self-attention for end-to-end speech
  recognition,''
\newblock in {\em Proc. Interspeech}, 2020.

\bibitem{zhou2020selfandmixed}
Xinyuan Zhou et~al.,
\newblock ``Self-and-mixed attention decoder with deep acoustic structure for
  transformer-based lvcsr,''
\newblock in {\em Proc. Interspeech}, 2020.

\end{thebibliography}

\end{document}